\documentclass[runningheads]{svmult}

\usepackage{makeidx}   
\usepackage{graphicx}  
\usepackage{subeqnar}  
\usepackage{multicol}  
\usepackage{physprbb}  
\makeindex             

\begin{document}
\title*{Weak Homology of Bright Elliptical Galaxies}
\toctitle{Weak Homology of Bright Elliptical Galaxies}
%
%
\titlerunning{Weak Homology of Bright Elliptical Galaxies}
%
\author{Giuseppe Bertin}
\authorrunning{G. Bertin}
%
%
\institute{Universit\`{a} degli Studi, Dipartimento di Fisica,
     via Celoria 16, I-20133 Milano, Italy}

\maketitle              

\begin{abstract}
Studies of the Fundamental Plane of early-type galaxies, from
small to intermediate redshifts, are often carried out under the
guiding principle that the Fundamental Plane reflects the
existence of an underlying mass-luminosity relation for such
galaxies, in a scenario where elliptical galaxies are homologous
systems in dynamical equilibrium. Here I will re-examine the issue
of whether empirical evidence supports the view that significant
systematic deviations from strict homology occur in the structure
and dynamics of bright elliptical galaxies. In addition, I will
discuss possible mechanisms of dynamical evolution for these
systems, in the light of some classical thermodynamical arguments
and of recent N-body simulations for stellar systems under the
influence of weak collisionality.
\end{abstract}

\section{Introduction}
This article focuses on three main questions: (1) Are elliptical
galaxies structurally similar to each other? (2) Which detailed
dynamical mechanisms can make elliptical galaxies evolve? (3) Are
there general trends to be anticipated for the evolution of these
stellar systems?

Here I will report on a long-term research project aimed at
providing answers to the above questions. Some interesting clues
have been discovered only very recently \cite{ber02a},
\cite{ber02b}, \cite{ber03}. Most of the paper refers to the class
of bright ellipticals only; low-luminosity ellipticals are known
to be characterized by different dynamical properties.

\section{Structure of bright elliptical galaxies}
The answer to whether elliptical galaxies can be considered to be
structurally similar to each other depends on the specific context
in which the question is posed and addressed. Below, I will focus
on the context of the physical interpretation of the Fundamental
Plane (\cite{dre87},  \cite{djo87}).

As demonstrated by a number of investigations (e.g., see
\cite{jor93}, \cite{jor96} for a study based on a sample of more
than 200 early-type galaxies), the observed correlation that
defines the Fundamental Plane, $\log{R_e} = \alpha \log{\sigma_0}
+ \beta SB_e + \gamma$ (with $\alpha = 1.25 \pm 0.1$, $\beta =
0.32 \pm 0.03$, $\gamma = - 8.895$ in the B band; the effective
radius being measured in $kpc$, the central velocity dispersion in
$km/sec$, the mean surface brightness in $mag/arcsec^2$
\cite{ben98}, \cite{jor93}), is remarkably tight, with a scatter
on the order of $15\%$ in $R_e$.

The following simple physical argument has been put forward as an
interpretation of this important physical scaling law. If we note
that (1) the observed luminosity law of bright elliptical galaxies
appears to be universal (the so-called $R^{1/4}$ law;
\cite{dev48}) and (2) the kinematical structure of these systems
is regular and uniform (\cite{ber94a}, \cite{ger01}), it is
natural to conclude that elliptical galaxies should be considered
as homologous dynamical systems, in the sense that the relevant
virial coefficient $K_V$ should be taken to be approximately
constant from galaxy to galaxy. Then, (3) given the existence of
the virial constraint, the Fundamental Plane can be seen as the
manifestation of a mass--luminosity relation for galaxies (see
\cite{fab87}, \cite{van95}). In fact, the virial theorem can be
written as $G M_{\star}/R_e = L (G/R_e)(M_{\star}/L) = K_V
\sigma_0^2$, where $M_{\star}$ is the mass of the luminous
component and $L$ is the total luminosity. By eliminating
$\sigma_0$ from the Fundamental Plane relation, one finds:

\begin{equation}
\left(\frac{M_{\star}}{L}\right)\frac{1}{K_V} \propto
R_e^{(2-10\beta + \alpha)/\alpha}L^{(5\beta - \alpha)/\alpha} \sim
L^{(5\beta - \alpha)/\alpha}. \label{1.1}
\end{equation}

\noindent The latter relation follows from the {\it empirical}
fact that $2-10\beta + \alpha \approx 0$.

\begin{figure}[b]
\begin{center}
\includegraphics[width=.6\textwidth]{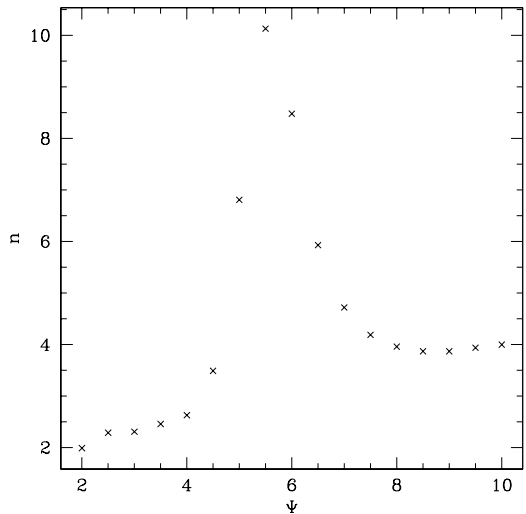}
\end{center}
\caption[]{The best-fit $n(\Psi)$, obtained by fitting the
$f_{\infty}$ models, projected along the line of sight, with
$R^{1/n}$ profiles. Note the plateau at $n = 4$ reached by
concentrated (high-$\Psi$) models, for which the radial range
adopted in the fit is $0.1 \leq R/R_e \leq 10$ (from
\cite{ber02a})} \label{eps1}
\end{figure}

Unfortunately, there are empirical and theoretical findings that
work against the hypotheses at the basis of the previous argument.
First of all, significant deviations from the $R^{1/4}$ law have
long been noted (see \cite{cao93}, \cite{pru97}), and found to
correlate systematically with the galaxy luminosity (see also
\cite{don94}). Second, studies that have measured the amount and
distribution of dark matter in ellipticals (see \cite{ber94a})
have shown that the presence of dark matter is more prominent in
brighter and spatially larger galaxies, thus demonstrating that
the virial coefficient may vary significantly from galaxy to
galaxy. A curious theoretical point adds further caution to the
perception that ellipticals should be considered homologous
systems. This derives from direct inspection of the so-called
$f_{\infty}$ sequence of models \cite{ber84}. As demonstrated in
\cite{ber02a}, models that appear to be all (see Fig.~1, for
$\Psi$ in the range $7 - 10$) very well fitted by the $R^{1/4}$
law, over a luminosity range of more than ten magnitudes, may be
characterized by significantly different values of the relevant
virial coefficient (see Fig.~2, the triangles representing the
virial coefficient for the $f_{\infty}$ sequence of models), as a
result of the impact of a more and more concentrated nucleus.

\begin{figure}[t]
\begin{center}
\includegraphics[width=.6\textwidth]{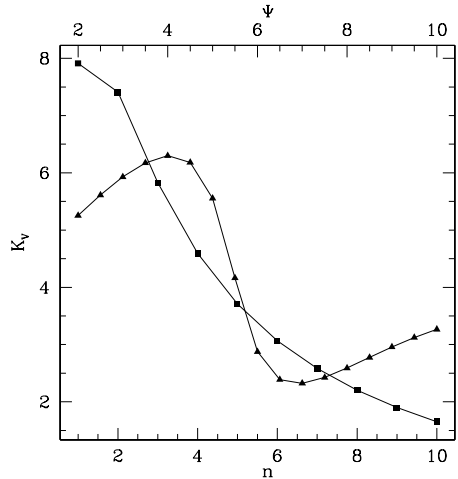}
\end{center}
\caption[]{The virial coefficient for the $f_{\infty}$ (triangles)
and for the isotropic $R^{1/n}$ (squares) models, based on an
aperture of radius $R_e/8$ (from \cite{ber02a})} \label{eps2}
\end{figure}

In \cite{ber02a} we have further confirmed, by close inspection of
four cases (NGC 1379, NGC 4458, NGC 4374, NGC 4552; studied in
great detail by comparing the performance of a number of fitting
procedures on data taken from \cite{cao90}, \cite{cao94}), that
the Sersic \cite{ser68} index $n$ for the $R^{1/n}$ photometric
profiles can indeed be very different from 4 (in particular, for
NGC 4552 we find $n \approx 11$, with residuals on the order of
$0.2$ magnitudes; a fit performed with the  $R^{1/4}$ law would
lead to residuals up to one magnitude, while a fit based on an
$R^{1/4} +$ exponential profile would have residuals up to half a
magnitude). On the other hand, we have checked that, if the
luminosity range where the fit to the photometric profile is
performed is reduced to less than 5 magnitudes, then (see
\cite{bur93}) the profiles are indeed well fitted by a
``universal" $R^{1/4}$ law.

In conclusion, while we find it necessary to dismiss strict
homology as a viable description of elliptical galaxies in
relation to the interpretation of the Fundamental Plane, the
existence of the empirical scaling law suggests that some kind of
{\it weak homology} must be enforced (expressed by
Eq.~(\ref{1.1})), as a correlation between structural properties
and total luminosity. In \cite{ber02a} we have also proved that a
large scatter in the dynamical correlations (e.g., in the $n \sim
- 19 + 3 \log{L}$ relation noted in \cite{cao93}, \cite{don94})
may well be compatible with the observed tightness of the
Fundamental Plane.

\section{Mechanisms of dynamical evolution}

Given the conclusion that elliptical galaxies have to be
considered only weakly homologous systems, it is natural to ask
whether and how individual galaxies may change their internal
structure via dynamical processes. This general issue is
especially important, if we recall that typically, in the study of
the cosmological evolution of the Fundamental Plane (see
\cite{tre02} and references therein), strict homology and thus a
mass--luminosity relation is assumed for the observed galaxies and
an interpretation of the data (see Fig.~3) is made in terms of
pure {\it passive evolution} (through the evolution of the
luminosity resulting from the evolution of the properties of
stellar populations).

\begin{figure}[ht]
\begin{center}
\includegraphics[width=.9\textwidth]{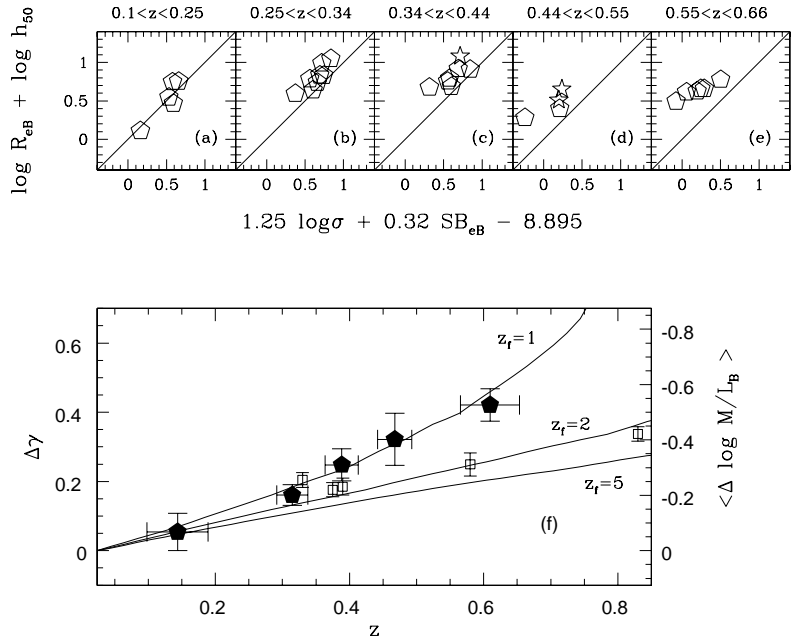}
\end{center}
\caption[]{The Fundamental Plane in the rest frame B band. In
panels (a) to (e), field E/S0 galaxies are shown, binned in
redshift, and compared to the Fundamental Plane found in the Coma
Cluster by \cite{ben98}. Panel (f) shows the average offset of the
intercept of field galaxies from the local Fundamental Plane
relation as a function of redshift (large filled pentagons)
compared to the offset observed in clusters (open squares). The
solid lines represent the evolution predicted for passively
evolving stellar populations formed in a single burst at $z = 1,
2, 5$ (from top to bottom). This figure is taken from \cite{tre02}
where full references are given to the sources for cluster data
points and stellar synthesis models} \label{eps3}
\end{figure}

Besides the possibility of major merger events, are there
significant sources of dynamical evolution for elliptical galaxies
to be considered? As noted recently \cite{ber00}, the traditional
approach to the study of elliptical galaxies, in terms of
equilibrium and stability for the solutions of the collisionless
Boltzmann equation, supplemented by the Poisson equation, may be
misinterpreted. Given the very large values of typical star-star
relaxation times in elliptical galaxies (see \cite{cha43},
\cite{spi87}) it is generally taken for granted that, unless a
system happens to be in a dynamically unstable state (for example,
a condition of excessive radial anisotropy; see \cite{pol81}), its
state is basically ``frozen" into an equilibrium distribution
function. Thus the only task left to the dynamicist would be to
decipher which distribution function best describes the observed
states (a task that is particularly difficult for non-spherical
systems) taken to be strictly stationary.

In our opinion, the above picture is oversimplified and may lead
to an improper perception of the dynamics of real stellar systems.
If, for simplicity, we take the view that elliptical galaxies have
formed via collisionless collapse (see \cite{van82}), we should
realize that splitting past and present conditions (that is
formation processes and a collisionless equilibrium state) is just
an idealization that the theory makes in order to define a basic
state and to study its properties. In reality, stellar systems
evolve continually and we should check to what extent the
evolution processes change the internal structure of galaxies.

There are several specific mechanisms and causes for dynamical
evolution that could be studied: (i) ``Granularity" in phase space
left over from the initial collapse; (ii) Presence of gas in
various phases, especially of the hot X-ray emitting interstellar
medium; (iii) Interactions with a compact central object; (iv)
Interactions between the galaxy and its own globular cluster
system; (v) Interactions with external satellites and effects of
tides and minor mergers.

In a recent paper \cite{ber02b} we have tried to quantify the role
of items (iv) and (v) above by means of N-body simulations. The
idea at the basis of these studies is that heavy objects can
suffer dynamical friction and then be dragged in toward the galaxy
center, as studied earlier, for example, in \cite{bon87},
\cite{bon88}, \cite{wei89}; in fact, the parallel momentum
transport relaxation time $T_{fr}$ is related to the deflection
relaxation time $T_D$ by a factor that can be very small when a
heavy test particle moves through a field of lighter particles:
$T_{fr} = 2 T_D m_f/(m_t + m_f)$. We have thus revisited the
problem of simulating the orbital decay of a satellite, placed
initially on a circular orbit at the periphery of a galaxy, and
basically confirmed the general findings presented in
\cite{bon88}; note that our simulations have been made with about
one million particles, while the earlier simulations had been
carried out with a few thousand particles. Then we have proceeded
to address a quasi-spherical problem in which the satellite is
fragmented into many smaller objects (several runs were made with
either 20 or 100 fragments), distributed on a spherical shell. The
quasi-spherical symmetry that characterizes this study has the
important advantage of allowing for a smoother framework,
basically free from other effects unrelated to dynamical friction,
such as those associated with lack of equilibrium in the initial
configuration. Furthermore, with respect to the earlier studies of
the orbital decay of a single satellite, our attention here is
mostly shifted to measuring the evolution of the underlying
structure of the hosting galaxies. One effect observed, while the
fragments are slowly dragged in toward the center, is a general
change in the stellar density distribution with respect to the
initial polytropic basic state. Another expected effect that we
have been able to quantify, starting from an initially isotropic
distribution of stellar orbits, is the slow growth of a
tangentially biased pressure anisotropy (see Fig.~4). All these
slow dynamical evolution effects appear to be genuinely associated
with the process of dynamical friction exerted by the stars on the
minority component of heavier objects. We are planning a survey of
cases that should allow us to identify general properties of
dynamical evolution in elliptical galaxies resulting from the
interaction between the stars and a significant population of
globular clusters or of the merging of a large number of small
satellites.

\begin{figure}[t]
\begin{center}
\includegraphics[width=.6\textwidth]{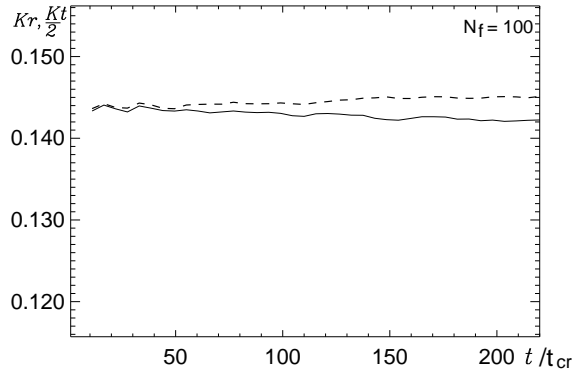}
\end{center}
\caption[]{The development of pressure anisotropy in a galaxy as a
result of the interaction with a shell of $N=100$ fragments
dragged in toward the galaxy center by dynamical friction. The
broken line represents the evolving value of $K_T/2$, where $K_T$
is the total kinetic energy associated with the star motions in
the tangential directions; the solid line represents the evolving
value of $K_r$, the total kinetic energy associated with the star
motions in the radial direction (from \cite{ber02b})} \label{eps4}
\end{figure}

\section{General trends from thermodynamical arguments}

In order to study possible general trends that may be anticipated
for the evolution of elliptical galaxies, we refer to the general
framework that has been successfully applied to the context of the
evolution of globular clusters. Globular clusters appear to be
well represented by King \cite{kin66} models (see \cite{djo94}).
They are recognized to be non-homologous stellar systems, subject
to dynamical evolution resulting from internal effects (such as
weak collisionality and evaporation) and external perturbations
(such as disk-shocking, when, in our Galaxy, their orbits happen
to cross the disk). It has been noted that these mechanisms of
dynamical evolution make a globular cluster evolve approximately
along the King equilibrium sequence (see \cite{ves97} and
references therein).

For globular clusters, an important paradigm is provided by the
{\it gravothermal catastrophe} \cite{lyn68}, which offers
interesting applications and physical interpretation (for a
review, see \cite{spi87}). Here we recall that, starting from the
study of isothermal gas spheres \cite{bon56}, the gravothermal
catastrophe is expected to occur also in stellar systems (see
\cite{ant62}, \cite{lyn68}). The instability is interpreted as due
to the curious property of self-gravitating systems of being
characterized by a negative effective specific heat. Although for
stellar systems a rigorous proof has been provided only for
idealized models where an isothermal set of stars is confined by a
spherical box, the paradigm is generally believed to be
sufficiently robust to be applicable to real stellar systems,
provided that they possess a sufficient level of internal
collisionality. An independent element that strengthens the view
that the paradigm is indeed robust has been added by an analysis
that has shown, for an isothermal gas, that spherical symmetry is
not a necessary ingredient \cite{lom01}.

Following some arguments initially put forward by Lynden-Bell (see
\cite{lyn67}, \cite{lyn68}), would there be a way to lay out a
similar scenario for elliptical galaxies as partially relaxed
stellar systems? If so, we would gather powerful
``thermodynamical" arguments to determine general trends for
evolution, beyond the specific paths produced by a given dynamical
mechanism.

In our view, there are two aspects of the problem that require
clarification. A first point is that we would like to start from a
physically justified equilibrium sequence, much like King models
for globular clusters, able to describe the general properties of
elliptical galaxies. A second point is that, formally, the origin
of the gravothermal catastrophe can be traced to the Poincar\'{e}
stability of linear series of equilibria (see \cite{kat78},
\cite{kat79}). For a proper mathematical derivation, one would
thus like to start from a sequence of collisionless models derived
rigorously from the Boltzmann entropy. In the absence of such a
sequence, a derivation of the gravothermal catastrophe has been
based on either an {\it unjustified ansatz} (see \cite{kat80},
\cite{mag98}), that the global temperature of the system would be
associated with the coefficient multiplying the energy in the
distribution function, or the use of non-standard entropies
\cite{cha02} (but for unrealistic models).

In order to address the first point, we may refer to a sequence of
models that have been found to be very promising for a realistic
description of elliptical galaxies (the so-called $f_{\infty}$
models; \cite{ber84}, see the review \cite{ber93}). These models
have been inspired by the characteristics of the products of
collisionless collapse, as derived from N-body simulations
\cite{van82}. In the simple spherical case, they are based on the
distribution function $f_{\infty} = A(- E)^{3/2}\exp{(-a E - c
J^2/2)}$, with $A, a, c$ positive constants, and define a
one-parameter equilibrium sequence, which, much like King models,
can be parameterized in terms of the dimensionless central
potential $\Psi = -a \Phi(0)$. For positive values of $E$ the
distribution function is taken to vanish.  When $\Psi$ increases
beyond a certain value, around $\Psi = 7$, the models have a
projected mass density profile that is well fitted by the
$R^{1/4}$ law and indeed they turn out to be an excellent tool to
fit the observations. From the point of view of statistical
mechanics, they have been found \cite{sti87} to be compatible with
a derivation based on a partition of phase space in terms of the
star energy and the star angular momentum square, under the
assumption that detailed conservation of the star angular momentum
is required at large values of angular momentum. This closely
follows our understanding of the process of partial violent
relaxation \cite{lyn67}. Unfortunately, the derivation is based on
heuristic arguments and the distribution function does not follow
from a straightforward exact mathematical extremization of the
Boltzmann entropy; in particular, the orbit time that acts as a
weight to the cells in phase space is replaced, for simplicity, by
a factor $1/(- E)^{3/2}$, which is approximately correct only for
weakly bound orbits. Therefore, attempts at using this equilibrium
sequence to study the gravothermal catastrophe in the context of
elliptical galaxies, while definitely appealing from the physical
point of view (see also \cite{ber88}), would remain less
satisfactory from the formal point of view.

Now we have shown \cite{ber03} that we can carry out a program
that is satisfactory not only from the physical point of view
(because it is based on an equilibrium sequence, also inspired by
studies of collisionless collapse \cite{van82}, that is able to
match the properties of observed galaxies), but also from the
mathematical point of view (because the distribution is derived
rigorously from the Boltzmann entropy by requiring the
conservation of a third global quantity $Q$, in addition to total
energy and total mass). The program is made possible by the second
option explored in \cite{sti87} for the construction of models of
partially relaxed stellar systems. This option leads to the
so-called $f^{(\nu)}$ models. It was already noted \cite{sti87}
that the general physical properties of the $f^{(\nu)}$ models are
close to those of the $f_{\infty}$ models and, in particular, that
for $\nu$ in the range $0.5 - 1$ their projected mass
distribution, for concentrated models, follows the $R^{1/4}$ law.

\begin{figure}[ht]
\begin{center}
\includegraphics[width=.6\textwidth]{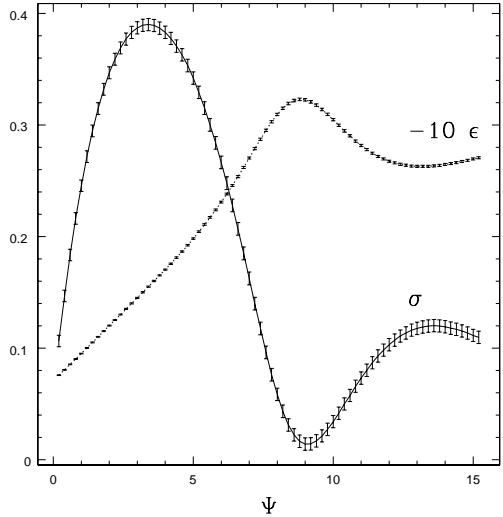}
\end{center}
\caption[]{Specific entropy and total energy along the equilibrium
sequence of $f^{(\nu)}$ models with $\nu = 1$ (as a function of
the concentration parameter $\Psi$, at constant $M$ and $Q$, and
thus expressed by means of the dimensionless functions
$\sigma(\Psi)$ and $\epsilon(\Psi)$). Note that for $\Psi < 3.5$
the models are characterized by a negative global temperature,
because the derivatives of $S$ and $E_{tot}$ have opposite signs.
This figure is taken from \cite{ber03}} \label{eps5}
\end{figure}

Let $f$ be the single-star distribution function, $E$ the
single-star specific energy, and $J$ the magnitude of the
single-star specific angular momentum. Consider the standard
Boltzmann entropy:

\begin{equation}
S = - \int f \ln{f} d^3v d^3x \label{1}
\end{equation}

\noindent and extremize it under the constraint that the total
mass

\begin{equation}
M = \int f d^3v d^3x, \label{2}
\end{equation}

\noindent the total energy

\begin{equation}
E_{tot} = \frac{1}{3} \int E f d^3v d^3x, \label{3}
\end{equation}

\noindent and a third global quantity

\begin{equation}
Q = \int J^{\nu}|E|^{-3\nu/4}f d^3v d^3x
\label{4}
\end{equation}

\noindent are assigned. Then the resulting distribution function
is

\begin{equation}
f^{(\nu)} = A \exp{(-a E - d J^{\nu}} |E|^{-3\nu/4}). \label{5}
\end{equation}

\noindent In the above expression, the quantities $A$, $a$, and
$d$ are positive constants. The parameter $\nu$ is a free
(positive) parameter, which was argued \cite{sti87} to be in the
range $0.5 - 1.0$. In the following we refer to the case $\nu =
1$. Note that the three constants appearing in the distribution
function define two scales and one dimensionless parameter, which
we take to be $\gamma = a d^{2/\nu}/(4 \pi G A)$.

Self-consistent models generated by such distribution function are
computed from the Poisson equation, solved under the boundary
conditions of regular potential at the center and of Keplerian
potential at very large radii. For positive values of $E$ the
distribution function is taken to vanish. If we introduce the
dimensionless central potential $\Psi = - a \Phi(0)$, the outer
boundary condition defines a sort of eigenvalue problem that is
solved by the relation $\gamma = \gamma(\Psi)$. The
self-consistent models thus make a one-parameter equilibrium
sequence.

\begin{figure}[t]
\begin{center}
\includegraphics[width=.6\textwidth]{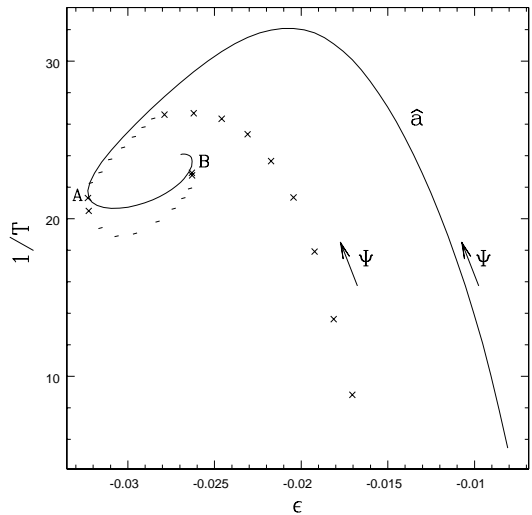}
\end{center}
\caption[]{The instability ``spiral" of $f^{(\nu)}$ models with
$\nu =1$. The solid line refers to the results obtained from the
{\it ansatz} that the coefficient $a$ represents the inverse
global temperature. Crosses represent the inverse global
temperature from the definition $\partial S/\partial E_{tot}$;
other symbols indicate estimated points for which the adopted
numerical differentiation is less reliable. Point A marks the
onset of the gravothermal catastrophe (from \cite{ber03})}
\label{eps6}
\end{figure}

By careful numerical integration, one may then proceed to
calculate the functions $S = S(M, Q, \Psi)$ and $E_{tot} = E_{tot}
(M, Q, \Psi)$ on the equilibrium sequence (see Fig.~5) and from
here the inverse {\it global temperature}

\begin{equation}
\zeta = \left(\frac{\partial S}{\partial E_{tot}}\right)_{M, Q}.
\label{6}
\end{equation}

\noindent The onset of the gravothermal catastrophe is thus
determined by inspection of the equilibrium sequence studied in
the $(E_{tot}, \zeta)$ plane (following \cite{kat78}).

In \cite{ber03} we have implemented the above program and shown
that for the $f^{(\nu)}$ models the gravothermal catastrophe is
expected to set in at $\Psi \approx 9$. Surprisingly, around this
value of the concentration, the projected mass distribution turns
out to be very well fitted by the $R^{1/4}$ law (this general
point had already been noted in \cite{sti87}, but outside the
context of the gravothermal catastrophe). For values of $\Psi$
close to and beyond 9, the general properties of the instability
``spiral" in the $(E_{tot}, \zeta)$ plane, based on the proper
thermodynamical definition of the global temperature, are the same
as in the $(E_{tot}, \hat{a})$ plane, based on the {\it ansatz}
that the temperature of the models is determined by the
coefficient $a$ (see Fig.~6).

One important point noted in \cite{ber03} is a qualitative
departure of the behavior of the instability ``spiral" at low
values of $\Psi$. For the original gas sphere and for the stellar
dynamical analogue of a stellar system confined by a box with
reflecting walls, the limit of low concentration was identified as
that of a {\it non-gravitating ideal gas}, subject to Boyle's law.
In our case, the analogy breaks down. In fact, the global
temperature turns out to {\it change sign} at $\Psi \approx 3.5$
(see Fig.~5). Such a drastic event should be accompanied by some
physical counterpart in the dynamical behavior of the system.
Surprisingly, the value of $\Psi \approx 3.5$ coincides with that
for the threshold of the radial-orbit instability \cite{pol81}
(for the context of $f_{\infty}$ models, see \cite{ber94b} and
\cite{ber89}). In other words, by undertaking a thermodynamical
description of the equilibrium sequence of models defined by the
$f^{(\nu)}$ distribution function, we have found arguments that
lead us naturally not only to the interpretation of the observed
$R^{1/4}$ law, but also to one clue for the interpretation of the
radial-orbit instability of collisionless stellar systems. Besides
the properties just outlined, one important additional aspect that
makes the $f^{(\nu)}$ models, at this point, more appealing than
the $f_{\infty}$ models is their anisotropy level. We had noted
(e.g., see \cite{ber93}) that the $f_{\infty}$ models are actually
too isotropic, when compared with the final products of
simulations of collisionless collapse \cite{van82}. The present
models turn out to be much more interesting even in this respect.
We have indeed checked that their characteristic anisotropy
profile, for values of $\Psi$ close to the onset of the
gravothermal catastrophe, is very similar to that observed in the
numerical simulations (see Fig.~7).

\begin{figure}[b]
\begin{center}
\includegraphics[width=.6\textwidth]{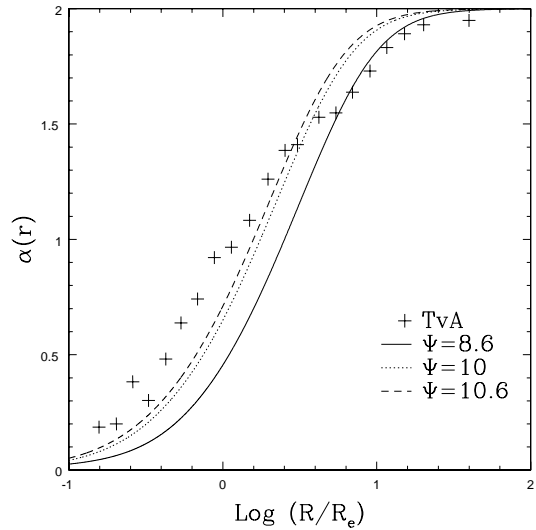}
\end{center}
\caption[]{Pressure anisotropy profiles $\alpha = 2 - (\langle
v^2_{\phi}\rangle + \langle v^2_{\theta}\rangle)/\langle
v^2_r\rangle$ as a function of radius for selected $f^{(\nu)}$
models ($\nu = 1$) compared to the pressure anisotropy profile
found \cite{van82} in numerical simulations of collisionless
collapse. This figure has been prepared by M. Trenti} \label{eps7}
\end{figure}

\section{Conclusions}

For a physically justified family of equilibrium models,
representing the result of incomplete violent relaxation, and
derived rigorously from the Boltzmann entropy, we have shown that,
at high concentration values, the onset of the gravothermal
catastrophe is found to occur at $\Psi \approx 9$, in the
parameter domain where models are characterized by an $R^{1/4}$
projected density distribution. At low concentration values, the
equilibrium sequence presents a drastic departure from the limit
of the classical isothermal sphere, because models become
associated with a negative global temperature. The transition
point, $\Psi \approx 3.5$, turns out to coincide with the point of
the sequence where the radial-orbit instability sets in. In the
intermediate concentration regime, $3.5 < \Psi < 9$, the
structural properties of the models change, much like those of
models along the King equilibrium sequence, a family of models
that is known to capture the non-homologous properties of globular
clusters. It is our hope that, in this domain of intermediate
concentration values, the $f^{(\nu)}$ models may be used to
describe the characteristics of weak homology of elliptical
galaxies.

\vspace{1cm}

\noindent {\it Acknowledgements}~~ I would like to thank L.
Ciotti, M. Del Principe, T. Liseikina, M. Lombardi, F. Pegoraro,
M. Stiavelli, M. Trenti, T. Treu, and T. van Albada, for their
collaboration in this work, and the organizers of the
workshop``Galaxies \& Chaos", for their warm hospitality in
Athens. This work has been partially supported by MIUR of Italy.

%

\end{document}